\documentclass[conference, 10pt]{IEEEtran}
\IEEEoverridecommandlockouts

\def\BibTeX{{\rm B\kern-.05em{\sc i\kern-.025em b}\kern-.08em
    T\kern-.1667em\lower.7ex\hbox{E}\kern-.125emX}}

\usepackage{cite}
\usepackage{amsmath,amssymb,amsfonts}
\usepackage{algorithmic}
\usepackage{graphicx}
\usepackage{textcomp}
\usepackage{xcolor}
\usepackage{makecell}

\usepackage{siunitx}
\usepackage{hyperref}
\usepackage[nolist]{acronym}
\usepackage[flushleft]{threeparttable}
\usepackage{placeins}
\usepackage{svg}                                   
\usepackage[percent]{overpic}
\usepackage[nameinlink, capitalise]{cleveref}

\usepackage{mathtools}
\usepackage{booktabs}
\usepackage[nolist]{acronym}
\usepackage{pbalance} 
\usepackage[export]{adjustbox}

\usepackage{eso-pic}    

\usepackage{orcidlink}

\usepackage{enumitem}

\usepackage{comment}

\usepackage{scalerel}

\usepackage{tabularx}
\usepackage{multirow}
\usepackage{wasysym}

\newcommand{\revise}[1]{{\textcolor{black}{#1}}}

\newcommand{\RNum}[1]{\uppercase\expandafter{\romannumeral #1\relax}}   

\DeclareSIUnit\db{dB}                           
\DeclareSIUnit\dbi{dBi}                         
\DeclareSIUnit\dbm{dBm}                         
\DeclareSIUnit\watthour{Wh}                     
\DeclareSIUnit\mbps{Mbps}                       
\DeclareSIUnit\kbps{kbps}                       
\DeclareSIUnit\bps{bps}                         
\DeclareSIUnit\msInference{ms/inference}        
\DeclareSIUnit\persquare{\ensuremath{/\square}}       

\newsavebox\racingcariconbox

\begin{document}
\bstctlcite{IEEEexample:BSTcontrol} 

\AddToShipoutPictureBG*{
  \AtPageUpperLeft{%
    \put(0,-40){\raisebox{15pt}{\makebox[\paperwidth]{\begin{minipage}{21cm}\centering
      \textcolor{gray}{This article has been accepted for publication in the proceedings of the \\
      International Workshop on Energy Harvesting \& Energy-Neutral Sensing Systems (Enssys 2026)\\ 
      } 
    \end{minipage}}}}%
  }
  \AtPageLowerLeft{%
    \raisebox{25pt}{\makebox[\paperwidth]{\begin{minipage}{21cm}\centering
      \textcolor{gray}{ \copyright 2025  Authors and IEEE. 
       This is the author’s version of the work. It is posted here for your personal use. Not for redistribution. \\
       The definitive Version of Record will be published in the proceedings of the International Workshop on Energy Harvesting \& Energy-Neutral Sensing Systems (Enssys 2026).
      }
    \end{minipage}}}%
  }
}


\title{A Class AAA Solar Testbed for Reproducible Long-Term Characterization of Energy-Harvesting Systems}

\author{
    \IEEEauthorblockN{
        Lukas Schulthess\,$^*$, 
        Andreas Rätz,
        Michele Magno,
        Philipp Mayer\,$^*$
    }
    \thanks{$^*$Corresponding authors: L. Schulthess (lukas.schulthess@pbl.ee.ethz.ch), P. Mayer (e-mail: mayerph@iis.ee.ethz.ch).}
    \vspace{1mm}
    \IEEEauthorblockA{
        \textit{Department of Information Technology and Electrical Engineering, ETH Zurich, Zurich, Switzerland}
    }
}

\maketitle

\begin{abstract}

Energy harvesting promises maintenance-free operation of wireless sensor nodes but introduces strong dependencies on stochastic and deployment-specific environmental conditions. In particular, solar-powered systems are highly sensitive to variations in irradiance and spectral composition, which complicates system-level design, parameter tuning, and reliable verification.

\revise{This work presents a solar testbed in which active control via \ac{hil} enables stable and repeatable illumination conditions for evaluating ultra-low-power energy harvesting systems.}
The proposed LED-based solar testbed provides spectrally configurable illumination over a wide dynamic range, from 5.7\,mW/m\textsuperscript{2} to 908\,kW/m\textsuperscript{2}. It achieves Class AAA performance according to IEC~60904-9, with a spectral match below 1.3\% and a spatial non-uniformity below 1.28\% over a 16.5\,cm$\,\times\,$16.5\,cm test area. The long-term irradiance instability remains below 0.6\%.
Closed-loop control using integrated illuminance and spectral sensors ensures high temporal stability, while a temperature-controlled DUT stage supports long-term experiments. Experimental results demonstrate high repeatability and suitability for systematic laboratory characterization of solar energy harvesting systems.

\end{abstract}

\vspace{5pt}

\section{Introduction}\label{sec:introduction}
Wireless sensor nodes enable the acquisition of data from physical and environmental phenomena in a wide range of industrial and cyber-physical applications, ranging from simple periodic sensing tasks~\cite{iot_sensing_moloudian_2024} to autonomous condition monitoring~\cite{eh_condition_monitoring_baszenski_2023} and predictive maintenance of industrial assets~\cite{eh_predictive_maintenance_2023}. A persistent challenge for such systems is the availability of a reliable long-term power source, particularly in large-scale deployments~\cite{eh_in_wsn_singh_2020, eh_for_wsn_avila_2025} and in scenarios with restricted physical access~\cite{passive_underwater_schuluka_2024}.

\begin{figure}
    \centering
    \includegraphics[width=\columnwidth]{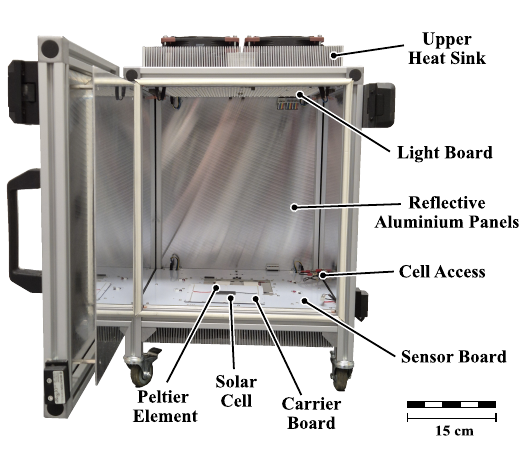}
    \vspace{-6mm}
    \caption{Hardware realization of the solar testbed. Reflective aluminum enclosure with a top-mounted, large-area, actively cooled light board providing spectrally configurable, high-precision irradiance control, and a bottom-mounted, temperature-controlled \ac{dut} platform with integrated illuminance and spectral sensors enabling closed-loop feedback; control and power electronics are mounted at the rear of the enclosure.}
    \vspace{-6mm}
    \label{fig:stb_overview}
\end{figure}

\ac{eh}, i.e., the exploitation of ambient energy sources to power electronic systems, has emerged as a promising approach to enable maintenance-free and potentially perpetual operation of wireless sensor nodes. While significant progress has been achieved in ultra-low-power circuit design, the practical realization of self-sustaining systems remains challenging.
Ambient energy sources are inherently intermittent and deployment-specific, and their stochastic nature must be carefully matched with the time-varying power demand of sensing, processing, and communication tasks~\cite{eco_wakeloc_cortesi_2026}. Consequently, system design requires an application-specific analysis of environmental energy statistics, combined with a detailed characterization of the power path, energy storage, and control algorithms, which substantially increases design and verification complexity~\cite{fast_inf_custode_2024, pzt_eh_peralta_2023, eh_modelling_mayer_2022, lp_eh_review_calautit_2021}.

Advanced energy management strategies and resilient power-path architectures, including cold-start-capable circuits~\cite{teg_eh_coldstart_lu_2024}, can partially mitigate these challenges by enabling recovery from temporary energy outages~\cite{battery_free_camera_giordano_2020}. However, guaranteeing reliable operation over the intended product lifetime remains difficult, particularly under challenging operating conditions~\cite{passive_underwater_schuluka_2024} or in the presence of rare but critical events~\cite{eh_for_event_warning_liu_2023}.
Short-term system-environment interactions, such as cold-start behavior, transient load conditions, or rapid variations in harvested power, are especially difficult to capture using purely analytical or simulation-based approaches~\cite{eh_simuation_challenges_mundu_2024}.

Consequently, experimental evaluation under controlled yet representative conditions becomes indispensable. Environmental testbeds enable systematic investigation of \ac{eh}-powered sensor nodes by emulating physical energy sources under well-defined conditions with high reproducibility and precision~\cite{testbed_concalves_2024, eh_testbed_sigrist_2021}. Such testbeds are essential for analyzing system-level behavior, validating design assumptions, and characterizing operational limits, while also enabling accelerated evaluation of long-term performance. In the absence of systematic evaluation and validation, designers are forced to adopt conservative design strategies with significant safety margins. As a result, energy harvesting systems often operate far from their optimal point, under-utilizing available environmental energy and obscuring the impact of dynamic system-environment interactions~\cite{pv_performance_assessment_ripathi_2025}.

This work addresses these challenges for solar-powered sensor nodes by introducing an \ac{hil} testbed that enables precise, repeatable control of illumination conditions (see \cref{fig:stb_overview}). Unlike conventional \ac{pv} characterization setups, the proposed platform is designed to support systematic evaluation of complete harvesting systems.
An LED-based light source provides software-configurable illumination spectra with high dynamic range, long-term stability, and excellent repeatability. This allows accurate reproduction of indoor and outdoor lighting conditions, ranging from ultra-low indoor illumination dominated by narrow-band artificial light spectra to full sunlight with a broad spectral distribution.
The main contributions of this work are summarized as follows:
\begin{enumerate}
    \item Design, implementation, and validation of a class AAA solar simulator, compliant with IEC~60904-9~\cite{norm_iec_60904_9_2020}, providing a uniform illumination area of \qty{16.5}{\centi\meter}$\times$\qty{16.5}{\centi\meter}. \revise{The hardware design of the system is open-sourced as part of this work\footnote{\url{https://github.com/ETH-PBL/Class-AAA-Solar-Testbed}}}.
    \item Spectrally tunable illumination capability enabling the emulation of both natural and artificial light sources over a dynamic irradiance range from 5.7\,mW/m\textsuperscript{2} to 908\,W/m\textsuperscript{2}, corresponding to illuminance levels from approximately \SI{1}{\lux} to \SI{110}{\kilo\lux}.
    \item A closed-loop illumination control architecture that continuously monitors and regulates the emitted light during experiments, enables reproducible harvesting measurements with high temporal stability, and reduces systematic errors in long-duration measurements.
\end{enumerate}

The remainder of this paper is organized as follows:
Section~\ref{sec:background} provides the relevant background and motivates the underlying standards for solar test equipment.
Section~\ref{sec:relatedWork} reviews related work and recent developments.
Section~\ref{sec:solar_testbed} describes the electrical and mechanical implementation of the proposed testbed.
Section~\ref{sec:results} presents the experimental results obtained using the testbed and discusses representative application scenarios.
Finally, Section~\ref{sec:conclusion} concludes the paper.

\section{Background}\label{sec:background}

\begin{table}[t]
    \centering
    \caption{Definition of solar simulator classifications \\ according to IEC 60904-9}
    \label{tbl:solar_simulator_classification}
    \setlength{\tabcolsep}{5pt}
    \renewcommand{\arraystretch}{1.2}
    \begin{tabularx}{\columnwidth}{
    >{\hsize=3\hsize\centering\arraybackslash}X
    >{\hsize=7\hsize\centering\arraybackslash}X
    >{\hsize=7\hsize\centering\arraybackslash}X
    >{\hsize=4\hsize\centering\arraybackslash}X
    >{\hsize=4\hsize\centering\arraybackslash}X
    }
        \toprule
        \makecell{\textbf{Class}} &
        \makecell{\textbf{Spectral} \\ \textbf{match}} &
        \makecell{\textbf{Spatial non-} \\ \textbf{uniformity (\%)}} &
        \makecell{\textbf{STI (\%)}} &
        \makecell{\textbf{LTI (\%)}} \\
        \midrule
        A & 0.75 -- 1.25 & 2  & 0.5 & 2  \\
        B & 0.6  -- 1.4  & 5  & 2   & 5  \\
        C & 0.4  -- 2.0  & 10 & 10  & 10 \\
        \bottomrule
    \end{tabularx}
    \vspace{-4mm}
\end{table}

Reliable evaluation of solar cells requires well-controlled, stable illumination conditions to ensure reproducible, comparable measurements~\cite{solar_cell_characterization_rauer_2024}. Consequently, standardized test methods and classification schemes for solar simulators have been established, primarily targeting the characterization of photovoltaic devices under well-defined irradiance and spectral conditions. In recent years, LED-based solar simulators have emerged as an attractive alternative to conventional lamp-based systems, offering high efficiency, long-term stability, and spectral controllability~\cite{led_solar_simulator_sun_2022}.

Although these standards were originally developed for solar cell testing, the same requirements are increasingly relevant for the evaluation of solar-based energy-harvesting systems and complete sensor nodes, particularly in ultra-low-power and self-sustaining IoT applications. In such systems, varying operating scenarios and deployment-specific illumination and spectral conditions must be accurately reproduced to enable reliable system-level analysis~\cite{solar_cell_varying_scenario_tonita_2025, available_energy_simulator_Geissdoerfer_2025}.

The importance of a reliable and well-controlled test environment is further underscored by IEC~60904-9~\cite{norm_iec_60904_9_2020} and ASTM~E927-19~\cite{ASTM_E927_19R25}. The former specifies performance requirements for solar simulators by defining criteria for spectral match, spatial non-uniformity, and temporal instability.
Each performance criterion is graded on a scale from A to C, defining the minimum requirements for each simulator class as summarized in \cref{tbl:solar_simulator_classification}. Temporal instability is further differentiated into \ac{sti} and \ac{lti} of the irradiance.
The spectral match of a solar simulator is defined by its deviation from the AM1.5G reference spectral irradiance, which represents the annual average solar irradiance at mid-latitude locations, as specified in IEC~60904-3~\cite{norm_iec_60904_3_2019}. For this purpose, the spectrum is divided into six wavelength intervals, listed in \cref{tbl:spectral_bins}. Within each spectral bin, the ratio of the simulator irradiance to the reference irradiance is evaluated and compared against the class-specific limits.

\begin{table}[b]
    \centering
    \vspace{-4mm}
    \caption{Wavelength intervals and corresponding irradiance contributions within the restricted range of 400\,nm to 1100\,nm}
    \label{tbl:spectral_bins}
    \setlength{\tabcolsep}{5pt}
    \renewcommand{\arraystretch}{1.2}
    \begin{tabularx}{\columnwidth}{
    >{\hsize=4\hsize\centering\arraybackslash}X
    >{\hsize=8\hsize\centering\arraybackslash}X
    >{\hsize=7\hsize\centering\arraybackslash}X
    >{\hsize=8\hsize\centering\arraybackslash}X
    }
        \toprule
        \textbf{Interval} &
        \textbf{Wavelength range (nm)} &
        \textbf{Irradiance fraction (\%)} &
        \textbf{Cumulative irradiance (\%)} \\
        \midrule
        1 & 400 -- 500  & 18.4 & 18.4 \\
        2 & 500 -- 600  & 19.9 & 38.3 \\
        3 & 600 -- 700  & 18.4 & 56.7 \\
        4 & 700 -- 800  & 14.9 & 71.6 \\
        5 & 800 -- 900  & 12.5 & 84.1 \\
        6 & 900 -- 1100 & 15.9 & 100.0 \\
        \bottomrule
    \end{tabularx}
\end{table}

\section{Related Work}\label{sec:relatedWork}
\begin{table*}[ht]
    \centering
    \caption{Comparison of related LED-based solar testbed systems}
    \label{tab:related_work}
    \setlength{\tabcolsep}{6pt}
    \renewcommand{\arraystretch}{0.5}
    \setcellgapes{3pt}
    \makegapedcells
    \begin{threeparttable}
    \begin{tabularx}{\textwidth}{
        >{\raggedright\arraybackslash}p{2.7cm}
        *{5}{>{\centering\arraybackslash}X}
    }
        \toprule
        &
        JPHOTOV'15~\cite{solar_simulator_novickovas_2015} &
        TIM'19~\cite{led_solar_simulator_lopez_2019} &
        SOLEN'22~\cite{led_solar_simulator_sun_2022} &
        G2V~\cite{g2voptics_sunbrick} &
        \textbf{This Work} \\
        \midrule

        \makecell[l]{Spectral Match\,\tnote{a}} &
        \makecell{$<7\%$ over \\ \qty{5}{\centi\meter} $\times$ \qty{5}{\centi\meter}}  &
        \makecell{$<3\%$ over \\ \qty{3}{\centi\meter} $\times$ \qty{3}{\centi\meter}} &
        \bfseries \makecell{\textless\,0.97\% over \\ 5\,cm $\times$ 5\,cm} &
        \makecell{$<2\%$ over \\ \qty{20}{\centi\meter} $\times$ \qty{20}{\centi\meter}} &
        \makecell{\textless\,1.35\% over \\ 16.5\,cm $\times$ 16.5\,cm} \\

        \makecell[l]{Spatial Non-uniformity} &
        \makecell{\textless\,2\% over \\ \qty{5}{\centi\meter} $\times$ \qty{5}{\centi\meter}} &
        \makecell{\textless\,1.77\% over \\ 1\,cm  $\times$ 1\,cm } & 
        \makecell{\textless\,1.5\% over \\ 5\,cm  $\times$ 5\,cm } & 
        \makecell{$<2\%$ over \\ \qty{20}{\centi\meter} $\times$ \qty{20}{\centi\meter}} &
        \bfseries\makecell{\textless\,1.28\% over \\  16.5\,cm $\times$ 16.5\,cm} \\

        \makecell[l]{Temporal Instability} &
        \makecell{$<0.25\%$ \\ over \qty{5}{\hour}} &
        \makecell{$<0.46\%$ \\ over \qty{5}{\minute}} &
        \makecell{\textless\,1\% \\ over 4\,h} & 
        \makecell{$<2\%$ \\ over \qty{1000}{\hour}} &
        \bfseries \makecell{\textless\,0.6\% \\ over 70\,h}  \\

        IEC Class\,\tnote{b} &
        \bfseries AAA &
        \bfseries AAA &
        \bfseries AAA &
        \bfseries AAA &
        \bfseries AAA \\

        \midrule

        Max DUT Size &
        6\,cm $\times$ 6\,cm &
        3\,cm $\times$ 3\,cm &
        7\,cm $\times$ 7\,cm &
        \bfseries 20\,cm\,$\times$\,20\,cm &
        16.5\,cm\,$\times$\,16.5\,cm \\

        Min Irradiance\,\tnote{c} &
        -- &
        -- &
        -- &
        100\,W/m\textsuperscript{2}&
        \bfseries 5.7\,mW/m\textsuperscript{2} \\
        
        Max Irradiance &
        \bfseries 1\,kW/m\textsuperscript{2} &
        818\,W/m\textsuperscript{2} &
        \bfseries 1\,kW/m\textsuperscript{2} &
        \bfseries 1\,kW/m\textsuperscript{2} &
         908\,W/m\textsuperscript{2} \\

        Spectral Controllability &
        No &
        \bfseries Yes &
        \bfseries Yes &
        \bfseries Yes &
        \bfseries Yes \\

        Spectral Sensing &
        No &
        No &
        No &
        No &
        \bfseries Yes \\

        Interface &
        Manual &
        LabVIEW &
        LabVIEW &
        Python, LabVIEW &
        Python \\

        \bottomrule
    \end{tabularx}

    \begin{tablenotes}
        \footnotesize
        \item[a] Worst-case deviation over all frequency bins. 
        \item[b] \ac{iec} class according to \ac{iec} 60904-9, based on spectral match, spatial non-uniformity, and temporal instability. 
        \item[c] The minimum achievable irradiance, which is non-zero.
    \end{tablenotes}
    \end{threeparttable}
    \vspace{-5mm}
\end{table*}

In line with the requirements introduced in the previous section, several LED-based solar simulators have been proposed with varying degrees of spectral precision, spatial uniformity, and controllability. 
Novickovas et al.~\cite{solar_simulator_novickovas_2015} presented a LED-based class AAA solar simulator designed to achieve accurate spectral matching and uniform illumination over a \qty{5}{\centi\meter} diameter area, with irradiance levels reaching up to 1\,kW/m\textsuperscript{2}. While the system satisfies grade~A classification criteria, its limited spectral tunability restricts its applicability primarily to standardized solar cell testing under fixed conditions.

Focusing on cost efficiency and modularity, L'opez-Fragu et al.~\cite{led_solar_simulator_lopez_2019} developed a low-cost class AAA system optimized for small photovoltaic devices. By combining 34 off-the-shelf \acp{led} covering wavelengths from \qtyrange{350}{1100}{\nano\meter}, the simulator closely approximates the AM1.5G spectrum. AAA classification is achieved within a central \qty{1}{\square\centi\meter} area, with a maximum irradiance of approximately \qty{818}{\watt\per\meter\squared}. Although the design emphasizes affordability and simplicity while maintaining acceptable spatial homogeneity and temporal stability, the usable high-accuracy area remains small.

Besides spectral matching and spatial uniformity, accurate reproduction of geometric illumination characteristics is essential when investigating a solar cell’s angular response. Sun et al. introduced an LED-based class AAA simulator incorporating a hyper-hemispherical aplanatic lens and a multi-source collimating system. The setup achieves one-sun AM1.5G irradiance with a beam divergence of approximately $\pm\,3^{\circ}$ and temporal instability below \qty{0.3}{\percent}, enabling precise reproduction of direct-sun conditions and reliable measurements of a solar cell’s angular response.

Finally, the commercial solar simulator \textit{G2V Optics Sunbrick} offers class AAA performance and a fully controllable spectrum~\cite{g2voptics_sunbrick}. The system supports an irradiance range from \qtyrange{12}{120}{\kilo\lux} and delivers uniform illumination over an area of \qty{20}{\centi\meter}$\,\times\,$\qty{20}{\centi\meter}.

Beyond purely photovoltaic simulators, Sigrist et al.~\cite{eh_testbed_sigrist_2021} presented a comprehensive testbed for evaluating low-power energy-harvesting systems that supports the emulation of photovoltaic and thermal sources. The platform enables accelerated testing through time scaling and detailed system-level characterization, but does not address standardized solar-simulator classification or spectral tunability.

In contrast to existing solutions, this work combines class AAA performance with a large test area of \qty{16.5}{\centi\meter}$\,\times\,$\qty{16.5}{\centi\meter}, a wide irradiance range from 10\,mW/m\textsuperscript{2} to 1\,kW/m\textsuperscript{2}, corresponding to illuminance levels from approximately \qty{1}{\lux} to \qty{120}{\kilo\lux}, and a fully controllable illumination spectrum.
Advanced optical sensing is integrated directly into the solar simulator’s illumination chamber, enabling closed-loop control of spectral composition and irradiance, as well as compensation for temperature-induced effects and long-term drift.
This combination enables precise and versatile characterization of photovoltaic and energy-harvesting systems across operating regimes ranging from ultra-low indoor illumination to full sunlight. In particular, it supports long-duration experiments under low-irradiance conditions relevant to indoor \ac{iot} applications, where available energy is limited and strongly influenced by deployment-specific spectral and temporal characteristics~\cite{indoor_solar_chakraborty_2024}.
A quantitative comparison with related work is summarized in \Cref{tab:related_work}.

\section{System Design}\label{sec:solar_testbed}
\begin{figure}
    \centering
    \includegraphics[width=\columnwidth]{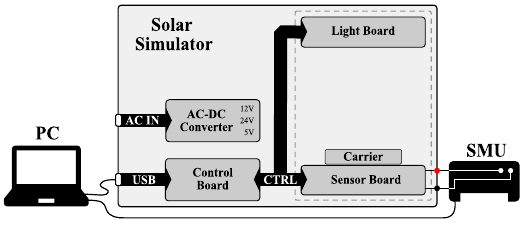}
    \vspace{-8mm}
    \caption{High-level block diagram of the solar testbed integrated into the measurement setup. The device under test is exemplarily connected to a Keysight B2902A precision source measurement unit for electrical characterization. The entire setup is controlled by a host computer, which orchestrates the measurements and logs the acquired data.}
    \vspace{-6mm}
    \label{fig:stb_high_level}
\end{figure}

The testbed has been designed as a versatile experimental platform that supports not only the characterization of photovoltaic devices but also their interaction with energy-harvesting circuitry and complete wireless sensor nodes. It is intended for integration into application-specific laboratory measurement setups, allowing combined evaluation with external, commercial measurement equipment, as illustrated in \cref{fig:stb_high_level}.
The system interfaces with a host computer via USB for configuration and data acquisition. The device under test is accessed through standard \qty{4}{\milli\meter} connectors, allowing either direct measurement of the photovoltaic response or connection to downstream energy-harvesting circuitry for system-level characterization.
The following subsections describe the electrical subsystems in detail.

\subsection{Electronics}
The testbed comprises four main subsystems, each implemented on a dedicated \ac{pcb} and responsible for illumination, sensing, carrier, and control. The overall system is powered from the \qty{230}{\volt} AC mains and uses dedicated AC-to-DC converters to generate multiple voltage rails, providing \qty{24}{\volt}, \qty{12}{\volt}, and \qty{5}{\volt} for the individual subsystems. The following subsections describe the key components of each circuit and their roles within the overall system architecture.

\subsubsection{Light Board}
\begin{figure}[b]
    \centering
    \includegraphics[width=\columnwidth]{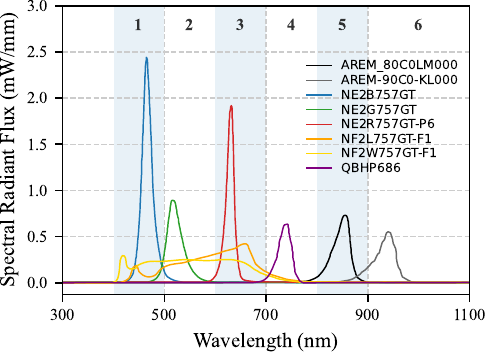}
    \caption{Overview of the selected \ac{led} emission spectra. The six wavelength intervals defined in IEC~60904-3 are highlighted in the background.}
    \label{fig:spectrum}
\end{figure}

The illumination architecture is based on a custom-designed light board enabling programmable spectral and intensity control.
It integrates a total of 2964 \acp{led}, comprising eight distinct \ac{led} types with unique emission spectra, as visualized in \cref{fig:spectrum} together with the six wavelength intervals defined in IEC~60904-3.
Two broadband \ac{led} types with high \ac{cri} cover the visible spectrum. To emulate natural illumination conditions, three infrared \ac{led} types with peak wavelengths at \qty{740}{\nano\meter}, \qty{850}{\nano\meter}, and \qty{940}{\nano\meter} extend the spectrum beyond the visible range. In addition, red, green, and blue \acp{led} enable fine-grained spectral adjustment.
The \acp{led} are driven by \textit{Renesas ISL78171} constant-current driver \acp{ic}, which implements a cascaded dimming scheme with a combined dimming resolution of \qty{16}{bit}. By independently adjusting the drive current of each \ac{led} type, the overall illumination spectrum can be precisely defined.
The light board further integrates \textit{Analog Devices MAX6615} fan-speed controllers to drive four fans, providing active cooling of the light source.

\subsubsection{Sensor Board}
The sensor board is integrated at the bottom of the testbed chamber and provides spectral, illuminance, and thermal monitoring, as well as control of the lower-stage actuators. It hosts an \textit{AMS AS7265x} 18-channel spectrometer array. In addition, two ambient light sensors with complementary sensitivity ranges are integrated to cover the full range of system illuminance. A \textit{Texas Instruments OPT3001} sensor provides high sensitivity at low illuminance levels, covering a range from \qty{0.01}{\lux} to \qty{83}{\kilo\lux}, while a \textit{Vishay VEML6031X00} sensor extends the measurable range up to \qty{228}{\kilo\lux} and beyond the visible spectrum.
The board further integrates four Peltier elements to enable temperature stabilization of the device under test. Thermal control is implemented using an \textit{Analog Devices AD5593} \ac{dac} in combination with four \textit{MPS MP8833GD-0000-Z} thermoelectric cooler controllers, which support both heating and cooling operations. Temperature stabilization during long-duration measurements is further improved by an \textit{Analog Devices MAX6615} fan-speed controller, which drives fans mounted on the lower heat sink.

\subsubsection{Carrier Board}
A dedicated \ac{dut} carrier board facilitates exchange, reliable fixation, and precise positioning of the \acl{dut} within the solar testbed.
\revise{It interfaces with the sensor board via an M.2 connector, which routes signals to banana-jack terminals for external access (see \cref{fig:stb_high_level}).
Mechanically secured above the Peltier elements, the setup enables active temperature regulation of the \ac{dut}.}
Temperature feedback is provided by an integrated \textit{Texas Instruments TMP116NAIDRVR} temperature sensor mounted on the carrier board.
The use of dedicated carrier boards improves experimental reproducibility by allowing different solar cells to be mounted and positioned at the exact same location within the illumination field, ensuring consistent measurement conditions across experiments.

\subsubsection{Control Board}
All subsystems described above are coordinated by the control board, which serves as the central interface for interacting with and operating the solar testbed. At its core sits an \textit{STMicroelectronics STM32L452VEI6} ARM Cortex-M4 \ac{mcu}, responsible for local control tasks including illumination spectrum adjustments, \ac{dut} temperature regulation, fan control, and system supervision. The microcontroller also provides external access via a serial interface over a USB Type-C connector.

\subsection{Control Interface}
For ease of use and integration into automated measurement setups, the solar testbed can be controlled via standardized \ac{scpi} commands over a serial connection. A dedicated Python module was developed, which abstracts the low-level command structure, enabling straightforward configuration, control, and data acquisition within experimental scripts, thereby improving usability and reproducibility.

\subsection{Mechanics}
The mechanical design of the solar testbed prioritizes safe and convenient operation in electronic laboratory environments. The structural frame is constructed from \qty{30}{\milli\meter}$\,\times\,$\qty{30}{\milli\meter} aluminum slot profiles and enclosed by reflective aluminum panels symmetrically arranged around the test area, fully encapsulating the \ac{dut}. This design ensures user safety by shielding against high-intensity illumination while simultaneously providing a controlled and disturbance-free measurement environment.
Operational safety is further enhanced by a front-access door equipped with a magnetic interlock that detects an open state and automatically disables the light source.
The internal dimensions of the enclosure are \qty{350}{\milli\meter}$\,\times\,$\qty{350}{\milli\meter}$\,\times\,$\qty{400}{\milli\meter}, allowing the integration of \acp{dut} with dimensions of up to \qty{210}{\milli\meter}$\,\times\,$\qty{210}{\milli\meter}.

Thermal management is realized using two custom aluminum heat sinks with finned structures designed to maximize convective heat dissipation. The upper heat sink is dedicated to cooling the \ac{led} board, while the lower heat sink maintains a stable operating temperature for the \ac{dut}. Each heat sink features fins with a height of \qty{37}{\milli\meter} and an exposed surface area of approximately \qty{3}{\meter\squared}. To further improve thermal stability during operation, active cooling is employed.

\section{Results}\label{sec:results}
To classify the presented solar testbed according to IEC~60904-9, the spectral match, spatial non-uniformity, and long-term temporal instability were evaluated.

\subsection{Spectral Match}
To evaluate the spectral match, the individual \acp{led} intensities were adjusted to accurately reproduce the AM1.5G spectrum, which serves as the reference spectrum for spectral match evaluation.
The individual spectral bins (see \cref{tbl:spectral_bins}) were characterized at six different irradiance levels ranging from 1\,W/m\textsuperscript{2} to 750\,W/m\textsuperscript{2}. Spectral measurements were performed using a \textit{Thorlabs CCS20} spectrometer with a spectral sensitivity range of \qtyrange{200}{1100}{\nano\meter}.
Comparing the achieved results listed in \cref{tab:avg_spectral_bins} with the IEC classification from \cref{tbl:solar_simulator_classification}, the solar testbed achieves class A for all tested irradiance levels.

\begin{table*}[t]
    \centering
    \caption{Average spectral-bin irradiance and distribution across different irradiance levels}
    \label{tab:avg_spectral_bins}
    \setlength{\tabcolsep}{5pt}
    \renewcommand{\arraystretch}{1.2}
    \begin{tabularx}{\textwidth}{
        >{\raggedright\arraybackslash}p{2.0cm}
        >{\centering\arraybackslash}p{2.0cm}
        *{6}{>{\centering\arraybackslash}X}
        >{\centering\arraybackslash}p{3cm}
    }
    \toprule
    \multirow{2}{*}{\makecell[c]{\textbf{Wavelength} \\ \textbf{Interval (nm)}}} &
    \multicolumn{6}{c}{\textbf{Intensity (\%) at Irradiance Level}} &
    \multirow{2}{*}{\textbf{Average (\%)}} &
    \multirow{2}{*}{\makecell[c]{\textbf{Irradiance Fraction} \\ \textbf{IEC 60904-9  (\%)}}} \\
    \cmidrule(lr){2-7}
    &
    1\,W/m$^{2}$ &
    10\,W/m$^{2}$ &
    50\,W/m$^{2}$ &
    300\,W/m$^{2}$ &
    500\,W/m$^{2}$ &
    750\,W/m$^{2}$ &
    \\
    \midrule
        400 -- 500   & 19.2 & 18.6 & 18.3 & 18.5 & 18.6 & 18.5 & 18.62 & 18.4\\
        500 -- 600   & 19.9 & 20.0 & 19.3 & 19.8 & 19.9 & 19.8 & 19.78 & 19.9\\
        600 -- 700   & 17.6 & 17.6 & 17.0 & 17.3 & 17.3 & 17.2 & 17.33 & 18.4\\
        700 -- 800   & 14.7 & 14.1 & 14.4 & 14.9 & 14.9 & 14.9 & 14.65 & 14.9\\
        800 -- 900   & 10.6 & 10.8 & 11.7 & 11.2 & 11.2 & 11.4 & 11.15 & 12.5\\
        900 -- 1100  & 17.9 & 18.7 & 19.1 & 18.2 & 18.0 & 18.2 & 18.35 & 15.9\\
        \bottomrule
    \end{tabularx}
    \vspace{-4mm}
\end{table*}

\subsection{Spatial Non-uniformity}
\begin{figure}[t]
    \centering
    \includegraphics[width=\linewidth]{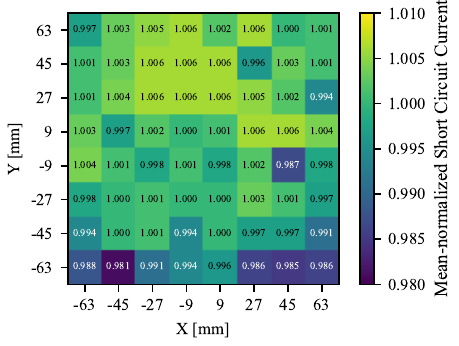}
    \vspace{-8mm}
    \caption{Mean-normalized spatial non-uniformity of the testbed measured at 500\,W/m\textsuperscript{2} under AM1.5G illumination using an \textit{Anysolar KXOB25-04X3F} solar cell.}
    \label{fig:spatial_non_uniformity}
    \vspace{-4mm}
\end{figure}

Spatial non-uniformity was evaluated by measuring the short-circuit current of an \textit{Anysolar KXOB25-04X3F} solar cell using a \textit{Keysight B2902A} precision \ac{smu}, which exhibits broad spectral sensitivity from \qty{300}{\nano\meter} to \qty{1100}{\nano\meter}.
The cell was translated across an 8$\,\times\,$8 measurement grid under constant irradiance, and the resulting short-circuit currents were recorded and normalized to the mean. The obtained spatial non-uniformity map is shown in \cref{fig:spatial_non_uniformity}.
Calculating the spatial non-uniformity yields a value of \qty{1.285}{\percent}, corresponding to class~A according to \cref{tbl:solar_simulator_classification}. However, the rear region of the light chamber (positive y) consistently exhibits a higher normalized short-circuit current than the front region (negative y), which contains the access door.
This indicates suboptimal reflective conditions near the door, resulting in a systematic irradiance gradient.

\subsection{Temporal Instability}
\begin{figure}[b]
    \centering
    \vspace{-4mm}
    \includegraphics[width=\linewidth]{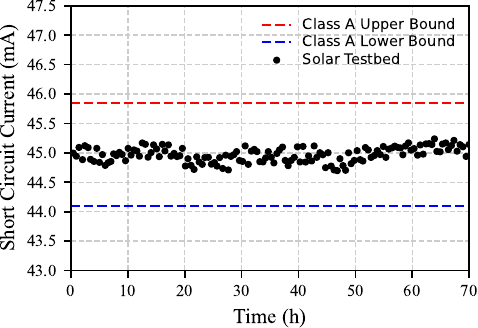}
    \vspace{-6mm}
    \caption{Measurements used for \ac{lti} calculation.}
    \label{fig:lti_temporal_stability}
\end{figure}

Short-term irradiance instability is defined in IEC~60904-9 for a single measurement interval, without specifying an explicit duration. For this evaluation, a measurement duration of \qty{3}{\minute} was selected. \revise{During this interval, the short-circuit current of an \textit{Anysolar KXOB25-04X3F} solar cell was measured at an irradiance level of 750\,W/m\textsuperscript{2} under AM1.5G conditions, while the cell temperature was actively stabilized using Peltier elements.}
\revise{Under these conditions, active feedback based on onboard spectrometer measurements was employed to continuously monitor irradiance and adjust LED intensity accordingly. This compensation of thermally induced drift improved the STI by approximately \qty{0.4}{\percent}, reducing it from \qty{0.43}{\percent} to \qty{0.24}{\percent}.}
Long-term irradiance instability was assessed by recording 140 short-circuit current measurements at an irradiance level of 500\,W/m\textsuperscript{2} under AM1.5G conditions. Consecutive measurements were taken at \qty{30}{\minute} intervals, resulting in a total evaluation period of \qty{70}{\hour}, as shown in \cref{fig:lti_temporal_stability}. Over this period, the long-term irradiance instability was within \qty{0.6}{\percent}. Both \ac{sti} and \ac{lti} satisfy the class~A requirements specified in IEC~60904-9.

\subsection{Dynamic Range}
To verify the achievable dynamic range of irradiance, a series of measurements was conducted to establish the quantitative relationship between the firmware-defined intensity settings for each of the eight \ac{led} types and their resulting irradiance output in W/m\textsuperscript{2}.
For this purpose, each \ac{led} was set to the lowest programmable intensity level of \qty{0.01}{\percent} and the highest programmable intensity level \qty{100}{\percent}.
The resulting irradiance was then measured using two instruments: a \textit{Kipp \& Zonen SMP10} pyranometer for higher irradiances, providing broadband measurements from \qty{270}{\nano\meter} to \qty{3000}{\nano\meter}, and a \textit{Gentec-EO PH100-SiUV-D0} photodiode detector for lower irradiances.
Summing the individual irradiance contributions of all \ac{led} types for both intensity levels yields a combined minimum irradiance of approximately 5.7\,mW/m\textsuperscript{2} and a maximum irradiance of 908\,W/m\textsuperscript{2}.
This irradiance range corresponds to illuminance levels from approximately \SI{1}{\lux} to \SI{110}{\kilo\lux}.

\begin{table}[ht]
    \centering
    \caption{Measured irradiance of the individual LED types at minimum (0.01\%) and maximum (100\%) intensity.}
    \label{tbl:led_irradiance}
    \setlength{\tabcolsep}{5pt}
    \renewcommand{\arraystretch}{1.2}
    \begin{tabularx}{\columnwidth}{
    >{\hsize=9\hsize\arraybackslash}X
    >{\hsize=4\hsize\centering\arraybackslash}X
    >{\hsize=7.5\hsize\centering\arraybackslash}X
    >{\hsize=6.5\hsize\centering\arraybackslash}X
    }
        \toprule
        \multirow{2}{*}{\textbf{LED Type}} &
        \textbf{Number of LEDs} &
        \textbf{Irradiance at 0.01\% (mW/m$^{2}$)} &
        \textbf{Irradiance at 100\% (W/m$^{2}$)} \\
        \midrule
        AREM-80C0-LM000     & 312   & 0.2501  & 74.111 \\
        AREM-90C0-KL000     & 648   & 0.4051  & 143.973 \\
        NE2B757GT           & 84    & 0.1267  & 38.376 \\
        NE2G757GT           & 72    & 0.3906  & 13.927 \\
        NE2R757GT-P6        & 120   & 0.4309  & 30.351 \\
        NF2L757GT-F1        & 420   & 3.0959  & 179.574 \\
        NF2W757GT-F1        & 780   & 0.2293  & 331.376 \\
        QBHP686             & 528   & 0.8382  & 97.266 \\
        \midrule
        \textbf{Total} & \textbf{2964} & \textbf{5.7668} & \textbf{908.914} \\
        \bottomrule
    \end{tabularx}
    \vspace{-4mm}
\end{table}

\section{Conclusion}\label{sec:conclusion}
This work presented a hardware-in-the-loop solar testbed for the systematic laboratory evaluation of ultra-low-power solar energy harvesting systems, combining a spectrally tunable LED-based light source with illumination control and an exchangeable, temperature-stabilized device-under-test stage.
Together, these elements enable precise, long-term experiments under well-defined illumination conditions, allowing reproducible characterization of complete energy-harvesting systems across indoor and outdoor scenarios.

A characterization according to the IEC~60904-9 specification demonstrated class~AAA performance with respect to spectral match, spatial non-uniformity, and temporal stability.
The ability to control irradiance over a wide range, from 5.7 mW/m\textsuperscript{2} to 908 W/m\textsuperscript{2}, enables realistic evaluation of energy-harvesting frontends under diverse lighting conditions, ranging from dim indoor environments dominated by narrow-band spectra of artificial light sources to full sunlight with a broadband spectrum. This is particularly relevant to \ac{iot} applications that target energy-autonomous operation.

Overall, the presented solar testbed provides a robust and versatile platform for reproducible characterization, validation, and optimization of solar-powered sensor nodes and energy harvesting systems. Decoupling system evaluation from stochastic environmental conditions, the proposed testbed enables reliable comparisons of designs and accelerates the development of energy-autonomous wireless sensing solutions.


\begin{acronym}

\acro{soa}[SoA]{State-of-the-Art}
\acro{hmi}[HMI]{Human-Machine Interface}
\acro{2d}[2D]{two-dimensional}
\acro{pc}[PC]{Personal Computer}
\acro{ee}[EE]{Electrical Engineering}
\acro{eh}[EH]{Energy Harvesting}
\acro{dut}[DUT]{Device under Test}
\acro{iec}[IEC]{International Electrotechnical Commission}
\acro{scpi}[SCPI]{Standard Commands for Programmable Instruments}

\acro{iot}[IoT]{Internet-of-Things}
\acro{fpga}[FPGA]{Field Programmable Gate Arrays}
\acro{cots}[COTS]{common off-the-shelf}
\acro{mcu}[MCU]{microcontrollers}
\acro{pll}[PLL]{phase-locked loop}
\acro{dma}[DMA]{Direct memory access}
\acro{opamp}[op-amp]{operational amplifiers}
\acro{fifo}[FIFO]{First In, First Out}
\acro{led}[LED]{light-emitting diode}
\acro{adc}[ADC]{analog-digital converter}
\acro{dac}[DAC]{Digital-to-Analog converter}

\acro{soc}[SoC]{system-on-chip}
\acro{ic}[IC]{integrated circuit}
\acro{rf}[RF]{radio frequency}
\acro{ble}[BLE]{Bluetooth low energy}
\acro{lut}[LUT]{Lookup Table}
\acro{tia}[TIA]{transimpedance amplifier}
\acro{wcet}[WCET]{Worst Case Execution Time}
\acro{uart}[UART]{Universal Asynchronous Receiver/Transmitter}
\acro{pwm}[PWM]{pulse-width modulation}
\acro{lptim}[LPTIM]{Low Power Timer}
\acro{fpu}[FPU]{Floating Point Unit}
\acro{gpio}[GPIO]{General-Purpose Input/Output}
\acro{simd}[SIMD]{Single Instruction Multiple Data}
\acro{i2c}[I2C]{InnInter-Integrated Circuit}
\acro{nvic}[NVIC]{Nested Vector Interrupt Controller}
\acro{spi}[SPI]{Serial Peripheral Interface}
\acro{sdo}[SDO]{Serial Data Out}
\acro{sdi}[SDI]{Serial Data IN}
\acro{sck}[SCK]{Serial Clock}
\acro{spst}[SPST]{Single-Pole Single-Throw}
\acro{pdm}[PDM]{Pulse Density Modulation}
\acro{dsp}[DSP]{Digital Signal Processing}
\acro{rtos}[RTOS]{Real-Time Operating Systems}
\acro{gpio}[GPIO]{General-Purpose Input-Output}
\acro{imu}[IMU]{Inertial Measurement Unit}
\acro{dma}[DMA]{Direct Memory Access}
\acro{cmsis}[CMSIS]{Common Microcontroller Software Interface Standard}
\acro{os}[OS]{operating system}
\acro{isa}[ISA]{Instruction Set Architecture}
\acro{cpu}[CPU]{Central Processing Unit}
\acro{pcb}[PCB]{Printed Circuit Board}

\acro{ppg}[PPG]{photoplethysmogram}

\acro{fft}[FFT]{Fast Fourier Transform}
\acro{snr}[SNR]{signal-to-noise ratio}
\acro{dsp}[DSP]{digital signal processing}
\acro{iir}[IIR]{infinite impulse response}

\acro{sd}[SD]{standard deviation}
\acro{rmse}[RMSE]{root mean square error}
\acro{mae}[MAE]{mean absolute error}
\acro{ae}[AE]{absolute error}
\acro{me}[ME]{mean error}

\acro{smu}[SMU]{Source Measurement Unit}

\acro{pbl}[PBL]{Project-Based Learning}

\acro{sti}[STI]{Short-term Instability}
\acro{lti}[LTI]{Long-term Instability}
\acro{cri}[CRI]{Colour Rendering Index}
\acro{hil}[HIL]{Hardware-in-the-Loop}
\acro{pv}[PV]{Photovoltaic}

\end{acronym}

\section*{ACKNOWLEDGMENT}
The authors would like to thank Florian Moll, Grischa Ruprecht, Romino Steiner, and Zun Hua for their valuable contributions and support during the development and experimental evaluation of the testbed.\\
This work was partially funded by the CHIST-ERA project ”SNOW” (Grant 209675).

\clearpage

\bibliographystyle{IEEEtranDOI} 

\bibliography{
    bib/sas
}

@IEEEtranBSTCTL{IEEEexample:BSTcontrol,
  CTLuse_forced_etal       = "yes",
  CTLmax_names_forced_etal = "3",
  CTLnames_show_etal       = "1",
  CTLuse_url               = "no",
  CTLuse_doi               = "no",
  CTLuse_eprint            = "no"
}

@standard{ASTM_E927_19R25,
  title        = {Standard Classification for Solar Simulators for Electrical Performance Testing of Photovoltaic Devices},
  organization = {ASTM International},
  number       = {E927-19R25},
  year         = {2025},
  volume       = {12.02},
  pages        = {8},
  doi          = {10.1520/E0927-19R25},
  note         = {Developed by Subcommittee E44.09; ICS Code: 27.160},
}

@manual{norm_iec_60904_9_2020,
     title         = "{IEC 60904-9:2020: Photovoltaic devices – Part 9: Solar simulator performance requirements}",
     shorttitle = {IEC 60904-9},
     author        = "International Electrotechnical Commission",
     organization  = "IEC",
     year          = "2020",
     month         = "December",
     url           = "https://webstore.iec.ch/publication/28973"
    }

@manual{norm_iec_60904_3_2019,
     title         = "{IEC 60904-3:2019: Photovoltaic devices – Part 3: Measurement principles for terrestrial photovoltaic (PV) solar devices with reference spectral irradiance data}",
    shorttitle = {IEC 60904-3},
     author        = "International Electrotechnical Commission",
     organization  = "IEC",
     year          = "2019",
     month         = "March",
     url           = "https://webstore.iec.ch/en/publication/64682"
    }

@article{led_solar_simulator_lopez_2019,
    author    = {Eduardo López-Fraguas and José M. Sánchez-Pena and Ricardo Vergaz},
    title     = {A Low-Cost LED-Based Solar Simulator},
    journal   = {IEEE Transactions on Instrumentation and Measurement},
    volume    = {68},
    number    = {12},
    pages     = {4913--4923},
    year      = {2019},
    doi       = {10.1109/TIM.2019.2899513},
}

@article{solar_simulator_novickovas_2015,
  author={Novičkovas, Algirdas and Baguckis, Artūras and Mekys, Algirdas and Tamošiūnas, Vincas},
  journal={IEEE Journal of Photovoltaics}, 
  title={Compact Light-Emitting Diode-Based AAA Class Solar Simulator: Design and Application Peculiarities}, 
  year={2015},
  volume={5},
  number={4},
  pages={1137-1142},
  doi={10.1109/JPHOTOV.2015.2430013}
}

@article{eh_testbed_sigrist_2021,
    author       = {L. Sigrist and A. Gomez and M. Leubin and J. Beutel and L. Thiele},
    title        = {Environment and Application Testbed for Low-Power Energy Harvesting System Design},
    journal      = {IEEE Transactions on Industrial Electronics},
    volume       = {68},
    number       = {11},
    pages        = {11146--11156},
    year         = {2021},
    doi={10.1109/TIE.2020.3036222}
}

@misc{g2voptics_sunbrick,
  title        = {{G2V Optics Sunbrick LED Solar Simulator}},
  author       = {{G2V Optics Inc.}},
  howpublished = {\url{https://g2voptics.com/products/led-solar-simulator-sunbrick}},
  note         = {Class AAA tunable LED solar simulator with modular, scalable illumination area and controllable spectrum},
  year         = {2025}
}

@inproceedings{passive_underwater_schuluka_2024,
  author={Schulthess, Lukas and Mayer, Philipp and Benini, Luca and Magno, Michele},
  booktitle={2024 International Symposium on Power Electronics, Electrical Drives, Automation and Motion (SPEEDAM)}, 
  title={A Passive and Asynchronous Wake-up Receiver for Acoustic Underwater Communication}, 
  year={2024},
  pages={480-485},
  doi={10.1109/SPEEDAM61530.2024.10609075}
}

@article{eh_condition_monitoring_baszenski_2023,
  title        = {Sensor integrating plain bearings: design of an energy-autonomous, temperature-based condition monitoring system},
  author       = {Baszenski, Thao and Kauth, Kevin and Kratz, Karl-Heinz and Guti\'errez Guzm\'an, Francisco and Jacobs, Georg and Gemmeke, Tobias},
  journal      = {Forschung im Ingenieurwesen},
  year         = {2023},
  volume       = {87},
  pages        = {441--452},
  doi          = {10.1007/s10010-023-00642-1},
}

@article{eh_predictive_maintenance_2023,
  title = {Development of an energy harvesting system based on a thermoelectric generator for use in online predictive maintenance systems of      industrial electric motors},
  author = {Luiz Fernando Pinto de Oliveira and Flávio José de Oliveira Morais and Leandro Tiago Manera},
  journal = {Sustainable Energy Technologies and Assessments},
  volume = {60},
  pages = {103572},
  year = {2023},
  doi = {10.1016/j.seta.2023.103572},
}

@article{eco_wakeloc_cortesi_2026,
  author={Cortesi, Silvano and Schulthess, Lukas and Plozza, Davide and Vogt, Christian and Magno, Michele},
  journal={IEEE Sensors Journal}, 
  title={Eco-WakeLoc: An Energy-Neutral and Cooperative UWB Real-Time Locating System}, 
  year={2026},
  volume={},
  number={},
  pages={1-1},
  doi={10.1109/JSEN.2026.3652283}
}

@article{eh_for_wsn_avila_2025,
  title = {Energy harvesting techniques for wireless sensor networks: A systematic literature review},
  journal = {Energy Strategy Reviews},
  volume = {57},
  pages = {101617},
  year = {2025},
  issn = {2211-467X},
  doi = {10.1016/j.esr.2024.101617},
  author = {Bernardo Yaser {León Ávila} and Carlos Alberto {García Vázquez} and Osmel {Pérez Baluja} and Daniel Tudor Cotfas and Petru Adrian Cotfas}
}

@article{eh_for_event_warning_liu_2023,
  title = {Ultra-high output hybrid nanogenerator for self-powered smart mariculture monitoring and warning system},
  journal = {Chemical Engineering Journal},
  volume = {472},
  pages = {145039},
  year = {2023},
  issn = {1385-8947},
  doi = {10.1016/j.cej.2023.145039},
  author = {Liqiang Liu and Jun Li and Zhengxin Guan and Leilei Zhao and Zhiyu Tian and Shouchuang Jia and Hongxin Hong and Zeyu He and Haiyang Wen and Ruiyuan Huang and Hui Cui and Wei Ou-Yang and Xiya Yang}
}

@article{eh_simuation_challenges_mundu_2024,
  title = {Simulation modeling for energy systems analysis: a critical review},
  author = {M. M. Mundu and S. N. Nnamchi and J. I. Sempewo and Daniel Ejim Uti},
  journal = {Energy Informatics},
  volume = {7},
  year = {2024},
  pages = {75},
  doi = {10.1186/s42162-024-00374-8},
}

@article{lp_eh_review_calautit_2021,
  title = {Low power energy harvesting systems: State of the art and future challenges},
  journal = {Renewable and Sustainable Energy Reviews},
  volume = {147},
  pages = {111230},
  year = {2021},
  issn = {1364-0321},
  doi = {10.1016/j.rser.2021.111230},
  author = {Katrina Calautit and Diana S.N.M. Nasir and Ben Richard Hughes}
}

@article{eh_in_wsn_singh_2020,
  title = {Energy harvesting in wireless sensor networks: A taxonomic survey},
  author = {Singh, Jaspreet and Kaur, Ranjit and Singh, Damanpreet},
  journal = {International Journal of Energy Research},
  year = {2020},
  volume = {45},
  number = {1},
  pages = {118--140},
  doi = {10.1002/er.5816},
}

@article{eh_modelling_mayer_2022,
  title = {Model-based design for self-sustainable sensor nodes},
  journal = {Energy Conversion and Management},
  volume = {272},
  pages = {116335},
  year = {2022},
  issn = {0196-8904},
  doi = {10.1016/j.enconman.2022.116335},
  author = {Philipp Mayer and Michele Magno and Luca Benini}
}

@ARTICLE{iot_sensing_moloudian_2024,
  author={Moloudian, Gholamhosein and Hosseinifard, Mohammadamin and Kumar, Sanjeev and Simorangkir, Roy B. V. B. and Buckley, John L. and Song, Chaoyun and Fantoni, Gualtiero and O’Flynn, Brendan},
  journal={IEEE Sensors Journal}, 
  title={RF Energy Harvesting Techniques for Battery-Less Wireless Sensing, Industry 4.0, and Internet of Things: A Review}, 
  year={2024},
  volume={24},
  number={5},
  pages={5732-5745},
  doi={10.1109/JSEN.2024.3352402}
}

@inproceedings{testbed_concalves_2024,
author = {Gon\c{c}alves, Diogo and Soares, Miguel Roque and Moreira, Waldir},
title = {LIGHTED: a Multi-energy Harvesting Testbed},
year = {2024},
isbn = {9798400712968},
publisher = {Association for Computing Machinery},
address = {New York, NY, USA},
doi = {10.1145/3698384.3699614},
booktitle = {Proceedings of the 12th International Workshop on Energy Harvesting and Energy-Neutral Sensing Systems},
pages = {48–49},
numpages = {2},
location = {Hangzhou, China},
series = {ENSsys '24}
}

@inproceedings{fast_inf_custode_2024,
author = {Custode, Leonardo Lucio and Farina, Pietro and Yildiz, Eren and Kilic, Renan Beran and Yildirim, Kasim Sinan and Iacca, Giovanni},
title = {Fast-Inf: Ultra-Fast Embedded Intelligence on the Batteryless Edge},
year = {2024},
isbn = {9798400706974},
booktitle = {Proceedings of the 22nd ACM Conference on Embedded Networked Sensor Systems},
publisher = {Association for Computing Machinery},
address = {New York, NY, USA},
doi = {10.1145/3666025.3699335},
pages = {239–252},
numpages = {14},
location = {Hangzhou, China},
series = {SenSys '24}
}

@article{solar_cell_characterization_rauer_2024,
  author = {Rauer, Michael and Fell, Andreas and Wöhler, Wilkin and Hinken, David and Reichel, Christian and Bothe, Karsten and Schubert, Martin C. and Hohl-Ebinger, Jochen},
  title = {The Impact of Measurement Conditions on Solar Cell Efficiency},
  journal = {Solar RRL},
  volume = {8},
  number = {3},
  pages = {2300873},
  doi = {10.1002/solr.202300873},
  year = {2024}
}

@article{led_solar_simulator_sun_2022,
  title = {LED-based solar simulator for terrestrial solar spectra and orientations},
  journal = {Solar Energy},
  volume = {233},
  pages = {96-110},
  year = {2022},
  issn = {0038-092X},
  doi = {10.1016/j.solener.2022.01.001},
  author = {Chao Sun and Zhiliang Jin and Yang Song and Yinhong Chen and Daxi Xiong and Kaiqiu Lan and Yang Huang and Mingliang Zhang}
}

@inbook{available_energy_simulator_Geissdoerfer_2025,
  author = {Geissdoerfer, Kai and Splitt, Ingmar and Sokolowski, Matthias and Herrmann, Carsten and Kubicki, Jonas and de Winkel, Jasper and Zimmerling, Marco},
  title = {Shepherd Nova: A Public Testbed for Rigorous Experiments Under Repeatable Energy-Harvesting Conditions},
  year = {2025},
  isbn = {9798400714535},
  publisher = {Association for Computing Machinery},
  address = {New York, NY, USA},
  booktitle = {Proceedings of the 23rd Annual International Conference on Mobile Systems, Applications and Services},
  pages = {236–248},
  doi = {10.1145/3711875.3729146}
}

@article{pzt_eh_peralta_2023,
title = {Design optimisation of piezoelectric energy harvesters for bridge infrastructure},
journal = {Mechanical Systems and Signal Processing},
volume = {205},
pages = {110823},
year = {2023},
issn = {0888-3270},
doi = {10.1016/j.ymssp.2023.110823},
author = {P. Peralta-Braz and M.M. Alamdari and R.O. Ruiz and E. Atroshchenko and M. Hassan},
}

@article{solar_cell_varying_scenario_tonita_2025,
  title = {A general illumination method to predict bifacial photovoltaic system performance},
  journal = {Joule},
  volume = {7},
  number = {1},
  pages = {5-12},
  year = {2023},
  issn = {2542-4351},
  doi = {10.1016/j.joule.2022.12.005},
  url = {https://www.sciencedirect.com/science/article/pii/S2542435122005761},
  author = {Erin M. Tonita and Christopher E. Valdivia and Annie C.J. Russell and Michael Martinez-Szewczyk and Mariana I. Bertoni and Karin Hinzer}
}

@article{pv_performance_assessment_ripathi_2025,
  title        = {Performance assessment of solar PV panels under varying environmental conditions: a laboratory and field-based approach for sustainable energy in mining operations},
  author       = {Tripathi, Abhishek Kumar and Aruna, Mangalpady and Sharma, Sumit and Kumar, Chandan and Didwania, Mukesh},
  journal      = {Environmental Science and Pollution Research},
  year         = {2025},
  publisher    = {Springer},
  doi          = {10.1007/s11356-025-35983-7},
  url          = {https://link.springer.com/article/10.1007/s11356-025-35983-7}
}

@article{indoor_solar_chakraborty_2024,
  title = {Photovoltaics for indoor energy harvesting},
  journal = {Nano Energy},
  volume = {128},
  pages = {109932},
  year = {2024},
  issn = {2211-2855},
  doi = {10.1016/j.nanoen.2024.109932},
  author = {Abhisek Chakraborty and Giulia Lucarelli and Jie Xu and Zeynab Skafi and Sergio Castro-Hermosa and A.B. Kaveramma and R. Geetha Balakrishna and Thomas M. Brown}
}

@ARTICLE{teg_eh_coldstart_lu_2024,
  author={Lu, Tianqi and Wang, Ruizhi and Tang, Zhong and Zou, Yiwei and Yue, Xinling and Liang, Yansong and Gong, Haoran and Liu, Shurui and Chen, Zhiyuan and Liu, Xun and Jiang, Junmin and Zhao, Bo and Du, Sijun},
  journal={IEEE Transactions on Power Electronics}, 
  title={A Thermoelectric Energy Harvesting System Assisted by a Piezoelectric Transducer Achieving 10-mV Cold-Startup and 82.7% Peak Efficiency}, 
  year={2024},
  volume={39},
  number={5},
  pages={6352-6363},
  doi={10.1109/TPEL.2024.3362366}
}

@inproceedings{battery_free_camera_giordano_2020,
  author = {Giordano, Marco and Mayer, Philipp and Magno, Michele},
  title = {A Battery-Free Long-Range Wireless Smart Camera for Face Detection},
  year = {2020},
  isbn = {9781450381291},
  publisher = {Association for Computing Machinery},
  address = {New York, NY, USA},
  url = {https://doi.org/10.1145/3417308.3430273},
  doi = {10.1145/3417308.3430273},
  booktitle = {Proceedings of the 8th International Workshop on Energy Harvesting and Energy-Neutral Sensing Systems},
  pages = {29–35},
  series = {ENSsys '20}
}

\end{document}